\documentclass[conference]{IEEEtran}
\IEEEoverridecommandlockouts
\usepackage{cite}
\usepackage{amsmath,amssymb,amsfonts}
\usepackage{algorithmic}
\usepackage{graphicx}
\usepackage{textcomp}
\usepackage{xcolor}
\def\BibTeX{{\rm B\kern-.05em{\sc i\kern-.025em b}\kern-.08em
    T\kern-.1667em\lower.7ex\hbox{E}\kern-.125emX}}

\usepackage{microtype}                 
\PassOptionsToPackage{warn}{textcomp}  
\usepackage{textcomp}                 
\usepackage{mathptmx}    
\usepackage{times}                    
         
\usepackage[draft=true]{minted}
\usepackage{tikz}
\usetikzlibrary{positioning}
\usepackage{url}
\usepackage{hyperref}
\hypersetup{hidelinks}

\usepackage[pscoord]{eso-pic}
\newcommand{\placetextbox}[3]{% \placetextbox{<horizontal pos>}{<vertical pos>}{<stuff>}
\setbox0=\hbox{#3}% Put <stuff> in a box
\AddToShipoutPictureFG*{% Add <stuff> to current page foreground
\put(\LenToUnit{#1\paperwidth},\LenToUnit{#2\paperheight}){\vtop{{\null}\makebox[0pt][c]{#3}}}%
}%
}%

\begin{document}

\title{
Audiovisual Database with 360$^\circ$ Video and Higher-Order Ambisonics Audio for Perception, Cognition, Behavior, and QoE Evaluation Research

\thanks{\IEEEauthorrefmark{4}A joint institution of the Friedrich-Alexander-Universit{\"a}t Erlangen-N{\"u}rnberg (FAU) and Fraunhofer Institute for Integrated Circuits (IIS). \\
*This research was funded by the Deutsche Forschungsgemeinschaft (DFG, German Research Foundation) project number 444832250 - SPP 2236.\\}
}

\author{\IEEEauthorblockN{{Thomas Robotham\IEEEauthorrefmark{4}, Ashutosh Singla\IEEEauthorrefmark{2}, Olli S. Rummukainen\IEEEauthorrefmark{4}, Alexander Raake\IEEEauthorrefmark{2}, Emanuël A. P. Habets\IEEEauthorrefmark{4}}}
\IEEEauthorblockA{{\IEEEauthorrefmark{4}\textit{International Audio Laboratories Erlangen.} 
Erlangen, Germany; thomas.robotham@audiolabs-erlangen.de}}
\IEEEauthorblockA{{\IEEEauthorrefmark{2}\textit{Audiovisual Technology Group, TU Ilmenau.} 
Ilmenau, Germany; ashutosh.singla@tu-ilmenau.de}}
}

\maketitle

\begin{abstract}
Research into multi-modal perception, human cognition, behavior, and attention can benefit from high-fidelity content that may recreate real-life-like scenes when rendered on head-mounted displays. Moreover, aspects of audiovisual perception, cognitive processes, and behavior may complement questionnaire-based Quality of Experience (QoE) evaluation of interactive virtual environments. 
Currently, there is a lack of high-quality open-source audiovisual databases that can be used to evaluate such aspects or systems capable of reproducing high-quality content. With this paper, we provide a publicly available audiovisual database consisting of twelve scenes capturing real-life nature and urban environments with a video resolution of 7680$\times$3840 at 60 frames-per-second and with 4$^{th}$-order Ambisonics audio. These 360$^\circ$ video sequences, with an average duration of 60 seconds, represent real-life settings for systematically evaluating various dimensions of uni-/multi-modal perception, cognition, behavior, and QoE. The paper provides details of the scene requirements, recording approach, and scene descriptions. The database provides high-quality reference material with a balanced focus on auditory and visual sensory information. The database will be continuously updated with additional scenes and further metadata such as human ratings and saliency information.
\end{abstract}

\placetextbox{0.5}{0.08}{\fbox{\parbox{\dimexpr\textwidth-2\fboxsep-2\fboxrule\relax}{\footnotesize \centering Accepted paper. \copyright  2022 IEEE. Personal use of this material is permitted. Permission from IEEE must be obtained for all other uses, in any current or future media, including reprinting/republishing this material for advertising or promotional purposes, creating new collective works, for resale or redistribution to servers or lists, or reuse of any copyrighted component of this work in other works.  DOI: \href{https://doi.org/10.1109/QoMEX55416.2022.9900893}{<10.1109/QoMEX55416.2022.9900893}}}}

\vspace{-5pt}
\begin{tikzpicture}[overlay, remember picture]
\path (current page.north) node (anchor) {};
\node [below=of anchor, yshift=15pt] {
Accepted paper at 2022 14th International Conference on Quality of Multimedia Experience (QoMEX)};
\end{tikzpicture}

\section{Introduction}
\label{sec:Introduction}

\maketitle

Omni-directional or 360$^\circ$ video content is a popular immersive media format. Using high-definition cameras and spherical microphone arrays, we can capture the visual and auditory scene at a `real-life-like' level of fidelity. This allows humans or other living entities to be accurately represented, avoiding aspects like the `uncanny valley'~\cite{mori.2012}, which may occur when using computer-generated imagery. All features of the acoustic environment can be directly captured with spherical microphone arrays as opposed to being computed using complex acoustic simulators. Real-life scenes and events can be captured, archived, and finally experienced at remote locations. Furthermore, 360$^\circ$ video content presented through head-mounted displays (HMDs) provides a systematic and reproducible way of studying human behavior and the associated cognitive processes and cross-modal perception. However, to conduct such research, the stimuli must represent system-related, semantic, and/or perceptual attributes that excite the evaluation criteria. For example, certain scene characteristics may be included that are interpreted as salient audiovisual information by a person exploring the omnidirectional content.

Few public databases are available that can be used to evaluate video codecs for high-end resolutions, particularly in virtual reality (VR) scenarios. Furthermore, many published databases are targeted towards aspects of video quality \cite{Dataset_1, Dataset_2, Dataset_4, Xu2017_YT, Nasrabadi2019Dataset} and consequently do not contain audio. While some databases do contain audio \cite{Duan2017IVQAD, Li2017valence, Chao2020_Ambi}, it is either non-spatialized or of a low spatial resolution. In terms of audio quality, databases containing spatial audio, for example, encoded using high-order Ambisonics (HOA)~\cite{frank.2015}, can increase spatial fidelity~\cite{Bertet.2013}. In turn, cross-modal effects and saliency may be investigated with a higher degree of auditory resolution. At a cognitive level, more holistic research including constituents of QoE such as media quality, simulator sickness and presence is becoming more popular in the context of 360$^\circ$ video for VR~\cite{Singla2019IEEEVR}. Here, longer exposure ($\geq$ 60~s) within `real-life-like' audiovisual scenes provides avenues of research with a degree of ecological validity more comparable to real life experiences than shorter duration's. Finally, many databases are derived from online video platforms such as YouTube. While this approach reduces production effort, little creative control is available, and finding the correct evaluation material can be time-consuming. 

This paper describes a database of twelve 360$^\circ$ video scenes that are targeted towards audio and visual quality, cognition, and behavior evaluation. For some scenes, multiple takes are available that offer different audiovisual interactions within the same setting. Video was captured using an omni-directional camera and post-processed files are provided at a high-resolution (7680$\times$3840). Audio was captured using a spherical microphone array capable of providing 4$^{th}$-order Ambisonics. The average duration of sequences is roughly 60~seconds, spanning different every-day scenes in nature and urban surroundings, featuring both specific events and contextual sources of interest.
The database is publicly available under the CC BY-NC 4.0 license and can be downloaded at \url{https://qoevave.github.io/database}.          

\section{Related Work}
\label{sec:RelatedWork}

\subsection{Audiovisual Content}
\label{sec:AudiovisualContent}

Recent research has explored the cross-modal influences of different auditory or visual formats. Chao et al. \cite{Chao2020_Ambi} use a database of 360$^{\circ}$ videos collected from YouTube to evaluate exploration for no, mono, and first-order Ambisonics (FOA) audio. Although salient auditory stimuli were shown to influence viewing behavior, the authors note that more research is needed to quantify cross-modality effects. Ferdig et al. \cite{ferdig.2020} reached similar conclusions regarding the effect of accompanying mono and FOA audio to 360$^{\circ}$ videos in the context of learning environments. Beyond FOA audio, Hirway et al. \cite{hirway.2020} conducted saliency evaluations with audiovisual content up to 3$^{rd}$-order Ambisonics using content available from \cite{farina.2022}. Their initial findings support the usefulness of Ambisonics audio with 360$^{\circ}$ video, suggesting better localization, identification, attention retainment, and enjoyment.

Audiovisual databases are also used for quality estimation. Fela et al. \cite{Fela.2020} evaluated various degradations through encoding parameters for video and audio data streams. While the paper focuses on audiovisual quality modeling via independent audio and visual mean opinion scores, the employed database~\cite{parma.2022} had a maximum resolution of 3840$\times$1920 at 29.97~frames-per-second (fps) and FOA audio. These technical limitations already limit the potential for certain audiovisual quality estimations using higher video resolution and Ambisonics orders. Audiovisual databases are further used to investigate perceptual attributes. Olko et al.~\cite{Olko.2017} authored audiovisual content of 2048$\times$1024 resolution video and FOA to identify perceptual attributes relevant for 360$^{\circ}$ VR. Kentgens et al.~\cite{kentgens.2018} establish a database of immersive content, capturing 3840$\times$1920 resolution video with 4$^{th}$-order Ambisonics to investigate perceptual attributes related to audiovisual three degrees-of-freedom (3-DoF) VR content. While both studies focus on auditory perception, there is little analysis that considers the impact on quality from cross-modal influences. In contrast, Rummukainen et al. \cite{Rummukainen.2014} investigated dimensions of scene categorization explicitly accounting for both audio and visual modality perception using a database of 19 audiovisual recordings of 5400$\times$2700 video resolution and FOA audio. Although providing some initial steps towards multi-modal perception in natural scenes, the authors conclude with remarks highlighting the need for further research to address the influences of individual modalities in an audiovisual setting.   

The demand for novel audiovisual content is increasing with more research emerging dedicated to understanding human exploration behavior and quality judgments in ecologically-valid, immersive settings. Nevertheless, most available databases are targeted toward uni-modal evaluation and their respective quality features. Therefore, it is important to understand the application and limitations of current studies not to overlook modality-specific requirements.

\subsection{360 Degree Video Content}
\label{sec:VideoContent}

Many 360$^{\circ}$ video databases have been published within research and standardization groups. This includes high-resolution 8192$\times$4096 (8K)  \cite{Dataset_1, Dataset_2, Liu2017_SJTU, Dataset_4}, and lower-resolution (4K) \cite{Duan2017IVQAD, Dataset_1, Dataset_2, Dataset_4, Zhang2017_IEEE, Mahmoudpour2019_Judder, Maugey2019Multiview360, Nasrabadi2019Dataset, Mazzola2021} video sequences. Xu et al.~\cite{xu.2020} provide a comprehensive review of 360$^{\circ}$ image, video, and saliency databases available for quality evaluation. These video databases are used in studies to evaluate the effect of various aspects of the processing pipeline throughout the acquisition, encoding, transmission, and consumption, regarding subjective test data and objective quality metrics starting from the highest-quality source possible. Aspects under study include stitching, geometrical map projection distortions, and compression artifacts such as mosquito noise, blocking and color bleeding, and spatial quality fluctuations with viewport movements. For extensive reviews on 360$^{\circ}$ visual quality features we refer the reader to~\cite{azevedo.2020, chirariotti.2021a}. In terms of the omnidirectional video scene contents, aspects such as high and low motion activity, and simple or complex spatial information~\cite{tran.2017} often stress difficulties of 360$^{\circ}$ video processing pipeline algorithms. Other aspects including varying levels of light exposure and color variation must also be represented in an evaluation database not to ignore non-360$^{\circ}$ video-related attributes~\cite{ITU_P910}.  

Recently, more research is conducted to understand the impact of 360$^{\circ}$ video parameters on cognitive processes within {VR}, including aspects such as simulator sickness and presence~\cite{Singla2019IEEEVR}. However, many of the available databases~\cite{Dataset_1, Duan2017IVQAD, Liu2017_SJTU, Mahmoudpour2019_Judder, Dataset_2, Dataset_4, Zhang2017_IEEE, Nasrabadi2019Dataset} intended for video quality are purposefully short. For aspects such as presence, it is uncertain if short duration sequences are sufficient to induce the necessary psychological states. Furthermore, longer duration sequences in combination with high visual fidelity may offer a degree of ecological validity when researching aspects representative of real-life experiences. 

\subsection{Spatial Audio Content}
\label{sec:AudioContent}

Ambisonics is a method of representing a sound field using spherical harmonics, where the Ambisonics order governs the spatial resolution \cite{frank.2015}. This three-dimensional description can be decoded into binaural playback and spatially rotated based on 3-DoF tracked head rotations within interactive content. For many 360$^{\circ}$ videos databases, audio channels are typically only mono or stereo as a by-product of the camera recording, or sound design and post-production efforts. While online libraries of Ambisonics recordings are available \cite{ambisonics.2022}, the contents are not curated strictly for the purposes of audio perception, cognition, and behavior evaluation. Weisser et al. \cite{Weisser.2019} provide a notable study which explicitly addresses this issue. Here, 13 scenes are recorded that are targeted towards realistic environments and subjects were asked to conduct a scene classification of the setting (indoor, outdoor - although no outdoor scenes are present, or combined) and rate `perceptual realism'. Particularly for QoE, high-quality spatial audio databases targeted to naturalistic settings are a rare but valuable addition to the research community. The most comparable contributions are soundscape studies, where spherical microphone arrays are used to capture the sonic atmosphere of the surrounding environment. For these databases \cite{Boren.2013, Aswathanarayana.2014, Roginska.2019, DeCoensel.2017}, spatial audio is only available in FOA. 

\subsection{Summary}
\label{sec:Summary}

Although individual databases are available that provide high-resolution videos or spatial audio, there are hardly any that combine these aspects and which place a balanced emphasis on the captured audiovisual content. As more research continues to explore perceptual, cognitive, and behavioral multi-modal aspects, there is a need for new high-quality 360$^{\circ}$ audiovisual material. Such a database can be used to research cross-modality interactions of real-life-like scenes in a systematic manner, in addition to targeting dimensions of independent audio and 360$^{\circ}$ video quality criteria.

\section{Database Requirements and Acquisition}
\label{sec:AudiovisualEvaluationDatabse}

\subsection{Requirements}
\label{sec:Requirements}

For capturing novel content that can be used over a broad range of audio, visual, and audiovisual research topics (Section~\ref{sec:RelatedWork}), we draw two fundamental requirements. $\mathcal{R}_1$: Fidelity of the captured content should be as close as possible to the real-life setting given current technological state-of-the-art limitations. Specifically, video information shall be stitched at the highest possible resolution (8K). Audio shall be captured using a spherical microphone array capable of delivering 4$^{th}$-order Ambisonics. $\mathfrak{R}_2$: The captured content shall represent real-life scenes that can be used to assess cognitive performances such as exploration and scene analysis, and be broad and varied enough to excite multiple dimensions of known perceptual aspects relevant for both auditory and visual quality evaluation. Furthermore, elements that are audio and visually correlated shall be targeted, fueling ongoing research into multi-modal perception and behavior.

\subsection{Content Capture}
\label{sec:ContentCapture}

The Insta360 Pro 2 camera was used to capture the video information of the scenes. The battery powered device has six F2.4 fisheye lenses that capture its surroundings with full 360$^\circ$ coverage. The resulting video from each lens is available in 4K resolution (3840$\times$1920) with a frame rate of 60~fps and encoded using H.264 with a bitrate of 120~Mbit/s. The camera height was adjusted to approximately 1.8~m, and all surfaces and objects of interest were kept at a distance of at least 2~m from the camera to avoid stitching and ghost artifacts when viewed through an HMD.

For audio capture, the em32 Eigenmike spherical microphone array with an outdoor windshield was employed. The Eigenmike can deliver up to 4$^{th}$-order Ambisonics and uses its own interface to connect with the recording device, in our setup, a laptop. For main power, a battery pack was built using a 12~v motorcycle battery. All devices (battery pack, interface, and laptop) were concealed during recording by positioning everything in the blind spot of the camera's south pole. The Eigenmike studio was used as recording software, and subsequently used to encode the 32 microphone signals into 4$^{th}$-order Ambisonics signals (i.e., $(4+1)^2=25$ audio signals), which are ordered using the Ambisonics channel numbering (ACN) and normalized using the Schmidt semi-normalization (SN3D).

\begin{center}
\begin{figure}[t]
\centerline{\includegraphics[width=\linewidth]{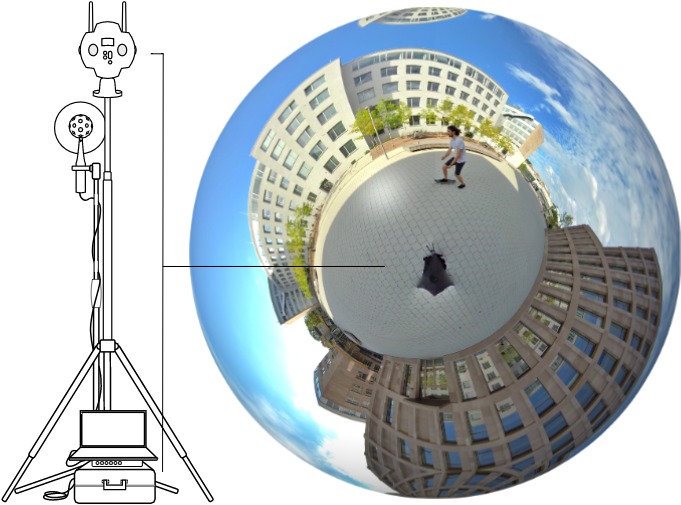}}
\caption{\textbf{Left}: Camera and microphone configuration. \textbf{Right}: Placement within the scene recording and top down blind-spot from camera.}
\label{fig:recordingSetup}
\end{figure}
\end{center}

The microphone was positioned as close to the camera stand as possible given the size of the windshield, with an elevation below the camera eye level not to obscure the visuals. It should be recognized that this offset yields perceivably incongruent spatial alignment with audiovisual sources only at very close distances. However, in all recordings, the intended audiovisual sources of interest were captured at a minimum distance of 2~meters. The complete recording setup is illustrated in Figure~\ref{fig:recordingSetup}. Before recording, both the camera and microphone were calibrated to the recording environment for visual stitching and noise levels, respectively. The rotation of the camera and microphone were aligned to face the same direction using the on-axis position of the devices to assist the audiovisual spatial alignment during post-editing. Upon recording, audiovisual signals were given with a clapperboard to later aid in audiovisual spatiotemporal alignment. 

\vspace{-2em}
\subsection{Post-Production}
\label{sec:Post-Editing}

The Insta360 Stitcher software\footnote{\url{https://www.insta360.com/de/download/insta360-pro} (v3.0.0).} was used to stitch the separate video recordings of the six individual lenses. Content type was selected as `Monoscopic' (for 2D 360° video) and stitching mode set to `optical flow', as most of the recorded scenes are complex in nature. For the sampling type, `fast' was selected as the camera was stationary in all of the scenarios. To ensure the highest quality of content, the software encoding speed was set to `highest' quality. Output settings of videos are selected as follows. The resolution, frame rate, and audio type of the video are selected to 7680$\times$3840, 60 fps, and `spatial audio' respectively (used to time-align the independent Eigenmike audio recordings). Finally, the output video is selected to be rendered in equirectangular projection (ERP) and encoded with H.265 in an MP4 container. 

Many tools are available in digital audio workstations (DAWs) for audio and video editing. However, the limitations of many DAWs is that they do not allow for high-resolution video editing beyond 2K (2048$\times$1080). Consequently, DAWs can be used as an intermediary step for finding start/stop timestamps, audio mixing, and editing, but video editing must be done separately to preserve video fidelity. For this workflow, the videos were cut to the desired duration with FFmpeg\footnote{https://ffmpeg.org/} using ffvhuff with the command:

\begin{minted}
[
frame=lines,
framesep=2mm,
baselinestretch=1.2,
fontsize=\footnotesize,
breaklines=true
]{bash}
ffmpeg -ss 00:00:15 -i Input_Video_Timestamp.mp4 -t 00:01:00 -c:v ffvhuff -pix_fmt yuv420p -y Output_Vid_Cut_60.mkv
\end{minted}

For audio editing, the Reaper DAW environment was used with several VSTs for Ambisonics processing such as the SPARTA \cite{McCormack.2019} or IEM plug-in suite\footnote{ https://plugins.iem.at}. Initially, wave files were cut to the timestamps of the desired start-/end-points and given a small 500ms fade-in/out to eliminate abrupt audio transitions. Afterward, the \textit{OmniCompression}, \textit{MultiEQ}, and \textit{SceneRotator}, plug-ins from the IEM suite were used for noise and alignment corrections. Firstly, the \textit{OmniCompression} was used to provide subtle make-up gain ($\leq$ +3~dB) to perceptually quiet recordings. Secondly, the \textit{MultiEQ} was used to apply a high-pass filter at 60~Hz to remove low-frequency vibrations, and a -3~dB notch filter at 10~kHz to reduce a more noticeable signal noise in scenes with a very low noise floor. Thirdly, the \textit{SceneRotator} was used to spatially align the audio to the video file if the audiovisual alignment was not correct. As many 360$^\circ$ video players do not allow for 8K video with 4$^{th}$-order Ambisonics and head tracking, the audio and video at this stage were played back separately in real-time using MaxMSP with the SPARTA AmbiBIN VST plug-in and Unity, respectively. For creating audiovisual content instances that can be previewed via YouTube, FOA audio was extracted from the 4$^{th}$-order Ambisonics files, muxed with the 8K video files into an MP4 container, and processed using Google's spatial metadata injector.

\section{Database Description and Applications}
\label{sec:Dataset}

\begin{figure*}[t]
\centering
\includegraphics[width=\linewidth]{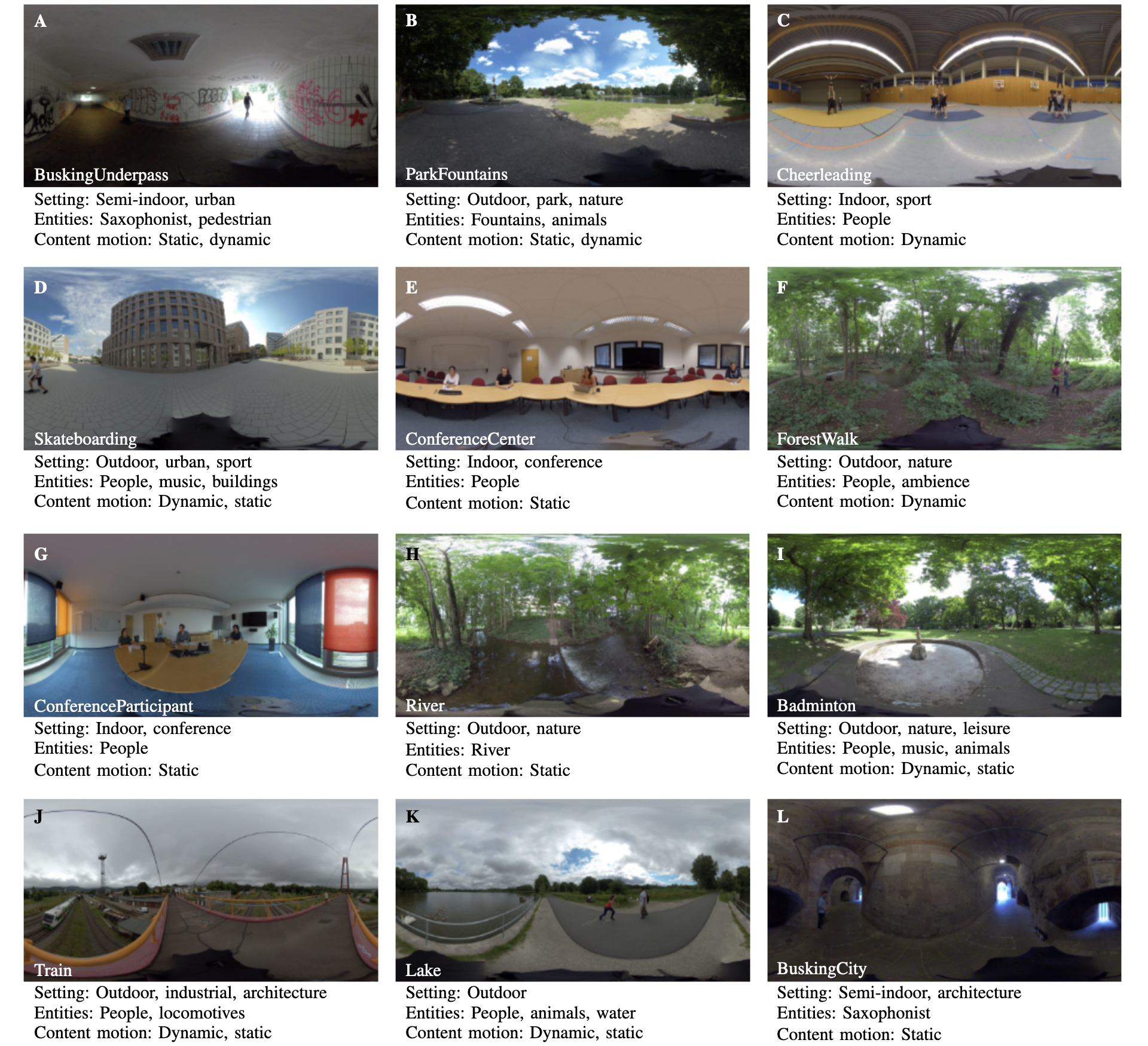}
\caption{Equirectangular frame captures for each recorded scene. Annotation labels include; \textbf{Settings}: environmental surroundings and gist. \textbf{Entities}: most prominent audio, visual, and audiovisual sources. \textbf{Content Motion}: static or dynamic movement throughout the scene recording.}
\vspace{-1.em}
\label{fig:SceneScreenshots}
\end{figure*}

All recordings took place in two main locations in Germany, namely across Ilmenau, Thuringia and Nuremberg, Bavaria. The recordings were shot during the day in indoor (3 scenes), semi-outdoor (2 scenes), and outdoor locations (7 scenes). Figure \ref{fig:SceneScreenshots} shows screenshots \textbf{A} through \textbf{L} of each location recording post-stitching. Additional meta information such as spatial and temporal information for the sequences were analyzed based on~\cite{ITU_P910} and are available at \url{https://qoevave.github.io/database}. The following sections describe some of the main attributes characterizing the recorded scenes that highlight example use-cases in uni-/multi-modal perception, cognition, and behavior evaluation research. 

\subsection{Multi-modal Perception}
\label{sec:ModalityPerception}

The resulting database possesses content which can be used to evaluate known dimensions of audio and video quality. For video quality, many of the sequences include a high degree of local motion from visual sources such as trains (\textbf{J}), and people (\textbf{A}, \textbf{B}, \textbf{D}, \textbf{F}) progressing from background to foreground positioning. These shifts in depth provide changes in both level of visual detail and viewing engagement throughout playback. As mentioned in Section~\ref{sec:RelatedWork}, video content with both high and low content motion is a valuable trait in 360$^\circ$ stimuli. The various levels of content motion in conjunction with fps settings may also influence perceived motion blur. It is noted that camera movements were excluded, as the focus of the database is on static observer positions. Other aspects include scenes with repeating textures like floor tiles (\textbf{A}, \textbf{D}), carpet patterns (\textbf{G}) and brick (\textbf{L}), highly detailed silhouetted shapes and colors from fauna (\textbf{F}, \textbf{H}, \textbf{I}), and also reflective surfaces such as water or windows (\textbf{D}, \textbf{H}, \textbf{K}), which all epmhasize areas related to high-resolution video encoding.     

For audio quality, the recorded source stimuli cover speech in reverberant, conference, and outdoor settings, music sources, trains, and varying natural sources prominent in urban / semi-urban soundscapes. In combination with the acoustic properties of the scene, multiple dimensions of perceptual auditory quality such as localization (all), speech clarity (\textbf{E}, \textbf{F}, \textbf{G}), distance perception (\textbf{A}, \textbf{D}, \textbf{F}, \textbf{J}), timbre and coloration (\textbf{A}, \textbf{D}, \textbf{F}, \textbf{L}), for example, are all excited within the provided database. Specifically, 4$^{th}$-order Ambisonics recording provides a level of spatial fidelity not available in many public databases, making the database a useful reference or training material for studies regarding auditory source separation and spatialization. 

As specified in the requirements, a main contribution of this database is the focus on the audiovisual cross-modal effects. Consequently, all scenes exhibit sources which are audiovisual such as instruments (\textbf{A} and \textbf{L} - saxophone), sport activities (\textbf{A} - skateboarding, \textbf{C} - cheerleading, \textbf{I} - badminton), trains (\textbf{J}), and speech in nearly all scenes. Each of these audiovisual sources varies in audio and visual properties providing a range of cross-modal interaction irrespective of the environmental properties. Examples are audiovisual source localization, or the visual texture and size of an instrument vs. its auditory timbre and tonality. 

Beyond the audiovisual sources themselves, environmental settings were selected to provide a further layer of cross-modal interaction. For example, the visual depth cues, setting, and material correspond to perceptual aspects of the reproduced sounds such as early reflections, reverberation, distance perception. In principle, many properties of audio and visual perception are intrinsically interrelated. A metal surface has a specific visual aesthetic and specular reflection which correlates with an acoustic absorption coefficient and the propagation of auditory reflections. From this perspective, the provided database encompasses different materials such as stone (\textbf{L}), tile (\textbf{A}), carpet (\textbf{E}, \textbf{G}), in additional to visual and acoustic environments such as free-field outdoor (\textbf{B}, \textbf{F}, \textbf{H}, \textbf{I}, \textbf{K}, \textbf{J}), diffuse-field indoor (\textbf{C}, \textbf{E}, \textbf{G}) and non-isotropic reverberant sound field (\textbf{A}, \textbf{D}, \textbf{L}) spaces. 
  
\subsection{Behavior and Cognition}
\label{sec:CognitionAndBehavior}

Within the twelve scenes, the behaviour of audio, visual, and audiovisual entities were purposefully scripted to achieve different viewing activity from subjects. This includes both static entities (\textbf{A}, \textbf{B}, \textbf{E}, \textbf{G}, \textbf{L}), dynamic/moving sources (\textbf{B}, \textbf{C}, \textbf{D}, \textbf{F}, \textbf{I}, \textbf{J}, \textbf{K}, \textbf{L}), and combinations thereof (\textbf{A}, \textbf{B}, \textbf{D}, \textbf{E}, \textbf{I}, \textbf{J}, \textbf{K}). The chosen entities also vary in their presence/absence of audio or visual properties. For example, a skateboard is a fast moving audiovisual entity (\textbf{D}), while some entities such as people moving do not have any complementary audio (\textbf{B}, \textbf{J}, \textbf{L}). Moreover, the spatial distribution of such activities also varies ranging from directional viewing of 110$^{\circ}$, to full 360$^{\circ}$ completely encompassing the viewer. Consequently, the provided database offers new content for evaluation in relevant areas of saliency-task-relevant analysis. Another main aspect of this contribution is the focus on real-life-like content within omnidirectional audio and video. Here, these scenes focus on placing a viewer as a by-stander or participant in an `ongoing world' to evaluate holistic aspects such as presence and plausibility in a setting more representative of `real-life'. Similar to \cite{Weisser.2019}, we provide content with naturalistic environments which allows a more ecologically valid representation of audiovisual interactions to assess how humans react and shift focus within real-life-like experiences.    

\section{Conclusion}
\label{sec:Conclusion}

In this paper, we present a new audiovisual database that can be used to evaluate multiple dimensions of perceptual audio, 360$^\circ$ video, and audiovisual quality. The initial release of the database contains twelve 360$^\circ$ scenes with 8K video and 4$^{th}$-order Ambisonics audio. For some scenes, multiple takes are provided that capture different audiovisual interactions in the same setting. The database was constructed to allow further research into high-order cognitive performances with an emphasis on real-life-like naturalistic settings for high audiovisual ecological validity. The database is publicly available at \url{https://qoevave.github.io/database}. In addition to human ratings and saliency data, the database will be continuously updated with further scenes targeted towards aspects such as simulator sickness and higher degrees-of-freedom while maintaining a balanced emphasis on audio and visual modalities.

\bibliographystyle{IEEEtran}
\bibliography{datasetBib,Video}
\end{document}